\documentclass[twocolumn,times,astrosymb]{aastex63111}

\usepackage{amsmath}

\newcommand{\older}{FRB~20181112A}
\newcommand{\newer}{FRB~20210912A}
\newcommand{\qp}{FRB~20230708A}
\newcommand{\utfrb}{FRB~20221128A}
\newcommand{\dg}{^{\circ}}
\newcommand{\pncr}{Poincar\'{e}}

\shorttitle{Intra-burst variations of polarization states in \newer\ and \qp\ }
\shortauthors{A. Bera et al.}
\graphicspath{{./}{figs/}}

\begin{document}

\title{Unusual intra-burst variations of polarization states in \newer\ and \qp\ : Effects of plasma birefringence?}

\author[0000-0002-2864-4110]{Apurba Bera}
\altaffiliation{E-mail: apurba.bera@curtin.edu.au}
\affiliation{International Centre for Radio Astronomy Research, Curtin University, Bentley, WA 6102, Australia}

\author[0000-0002-6437-6176]{Clancy~W.~James}
\altaffiliation{E-mail: clancy.james@curtin.edu.au}
\affiliation{International Centre for Radio Astronomy Research, Curtin University, Bentley, WA 6102, Australia}

\author[0009-0009-7038-1295]{Mark~M.~McKinnon}
\altaffiliation{E-mail: mmckinno@nrao.edu}
\affiliation{National Radio Astronomy Observatory, Socorro, NM 87801, USA}

\author[0000-0002-3532-9928]{Ronald~D.~Ekers}
\affiliation{Australia Telescope National Facility, CSIRO, Space and Astronomy, PO Box 76, Epping, NSW 1710, Australia}
\affiliation{International Centre for Radio Astronomy Research, Curtin University, Bentley, WA 6102, Australia}

\author[0009-0004-1205-8805]{Tyson~Dial}
\affiliation{Centre for Astrophysics and Supercomputing, Swinburne University of Technology, Hawthorn, VIC 3122 Australia}

\author[0000-0001-9434-3837]{Adam~T.~Deller}
\affiliation{Centre for Astrophysics and Supercomputing, Swinburne University of Technology, Hawthorn, VIC 3122 Australia}
\affiliation{OzGrav, the ARC Centre of Excellence for Gravitational Wave Discovery}

\author[0000-0003-2149-0363]{Keith~W.~Bannister}
\affiliation{Australia Telescope National Facility, CSIRO, Space and Astronomy, PO Box 76, Epping, NSW 1710, Australia}
\affiliation{Sydney Institute for Astronomy, School of Physics, The University of Sydney,NSW 2006,Australia}

\author[0000-0002-5067-8894]{Marcin~Glowacki}
\affiliation{International Centre for Radio Astronomy Research, Curtin University, Bentley, WA 6102, Australia}

\author[0000-0002-7285-6348]{Ryan~M.~Shannon}
\affiliation{Centre for Astrophysics and Supercomputing, Swinburne University of Technology, Hawthorn, VIC 3122 Australia}

\begin{abstract}

Fast radio bursts (FRBs) are highly energetic events of short-duration intense radio emission, the origin of which remains elusive till date. Polarization of the FRB signals carry information about the emission source as well as the magneto-ionic media the signal passes through before reaching terrestrial radio telescopes. Currently known FRBs show a diverse range of polarization, sometimes with complex features, making it challenging to describe them in a unified model.  \qp\ and \newer\ are two bright and highly polarized (apparently) one-off FRBs detected in the Commensal Real-time ASKAP Fast Transients (CRAFT) survey with the Australian Square Kilometre Array Pathfinder (ASKAP) that exhibit time-dependent conversion between linear and circular polarizations as well as intra-burst (apparent) variation of Faraday rotation measure. We investigate the intra-burst temporal evolution of the polarization state of radio emission in these two events using the \pncr\ sphere representation and find that the trajectories of the polarization state are well described by great circles on the \pncr\ sphere. These polarization features may be signatures of a transition between two partially coherent orthogonal polarization modes or propagation through a birefringent medium. We find that the observed variations of the polarization states of these two FRBs are qualitatively consistent with a magnetospheric origin of the bursts and the effects of propagation through a birefringent medium with linearly polarized modes located close to the emission source --- likely in the outer magnetosphere or near-wind region of a neutron star.
\end{abstract}


\section{Introduction} \label{sec:intro}

Fast radio bursts (FRBs) are impulsive events of radio emission reaching intensities so high that they are detectable from galaxies as distant as redshift $\sim 1$ \citep[e.g.][]{Lorimer2007,thornton2013,petroff22review,Ryder2023}. Despite the large number of FRBs discovered to date --- a small fraction of them known to repeat --- the source of these energetic bursts remains elusive and the emission mechanism is yet to be understood \citep[][]{chime21cat1,chime23repeater,zhang23review}. Several progenitor models have been proposed for both repeating and (apparently) non-repeating FRBs --- many of them involving neutron stars \citep[see e.g.][]{petroff22review} --- but no consensus has yet been reached \citep[see also][]{wang24ns}. FRBs exhibit vast diversity in their burst morphology and polarization characteristics \citep[e.g.][]{Day20askapol,CHIME_morphology_2021,Sherman2023,pandhi24chimepol}. These properties carry signatures of their emission source, as well as the magneto-ionic media along the line of sight. The effects of propagation through the interstellar media (ISM) --- in both the host galaxy of the FRB and the Milky Way --- and the intergalactic medium (IGM) are relatively better understood as they are not expected to change at the burst time scales ($\sim$ ms). However, the dense and complex magneto-ionic medium in the near vicinity of the source, which may change at shorter time-scales, is generally difficult to model \citep[e.g.][]{chen11mnras,ilie19rm,lyutikov22apj}. Therefore, high-time-resolution studies, which resolve polarization properties on millisecond or shorter timescales, are crucial for understanding the FRB source and emission mechanism.

A large fraction ($\gtrsim 60\%$) of the known FRBs are highly polarized; while the vast majority show predominantly linear polarization, a significant degree of circular polarization has been observed in $\sim$20\% of FRBs  \citep[e.g.][]{Sherman2023}. Some FRBs exhibit intra-burst temporal variation of the polarization state of radio emission \citep[e.g.][]{cho20apjl,Sherman2023}; the reason behind this is still not well understood. In this work, we investigate the intra-burst short timescale ($\sim$ 10 $\mu$s) variation of polarization state in two ASKAP-detected (apparently) one-off FRBs --- \qp\ \citep{dial25qp,shannon24craftics} and \newer\ \citep{Marnoch2023,twins} --- which show significant and varying degrees of linear and circular polarizations, as well as ``apparent'' variation of Faraday Rotation measure\footnote{Throughout this work we use ``apparent RM"/``observed RM" to refer to the slope of the observed position angle ($\rm PA_{obs}$) vs $\lambda^2$ curve. This quantity may not always be physically associated with Faraday rotation; when the PA has a different dependence on $\lambda$ \citep[e.g.][]{kennet98gfr,lyutikov22apj} the apparent RM only represents the local slope of the relation within the observed frequency band.}, across their burst profiles. The \pncr\ sphere representation provides a convenient way for analysis and interpretation of such variations of the polarization state of electromagnetic (EM) waves \citep[e.g.][]{rad94pol,malykin97pspreview}. We use the \pncr\ sphere representation to analyze the temporal evolution of the polarization vectors of \qp\ and \newer, and discuss possible interpretations for the observed variations of the polarization states. 

\vspace{-1em}
\begin{deluxetable}{lcc}
\tablecaption{\textbf{Properties of \newer\ and \qp\ } \label{tab:frbs}}
\tablewidth{0pt}
\tablehead{
\nocolhead{Properties} & \colhead{\newer\ } & \colhead{\qp\ } 
}
\startdata
J2000 RA & 23h23m10.35s & 20h12m27.73s\\
J2000 DEC & -30$\dg$24$\arcmin$19.2$\arcsec$ & -55$\dg$21$\arcmin$22.6$\arcsec$ \\
Host galaxy redshift & $\gtrsim 0.7$ & 0.105 \\
Central frequency $^{\dagger}$ & 1271.5 MHz & 919.5 MHz \\
DM$^{\ast}$ (pc cm$^{-3}$) & $1233.696 \pm 0.006$ & $411.51 \pm 0.05$ \\
Burst fluence (Jy\,ms) & $70 \pm 2$ & $111 \pm 4$ \\
RM$_{\rm avg}^{**}$ (rad m$^{-2}$) & 4.55 & -5.75 \\
\enddata
\tablecomments{
$^{\dagger}$ Centre of the 336 MHz observing band \\
$^{\ast}$ Structure maximizing dispersion measure \citep{Sutinjo2023} \\
$^{**}$ Faraday rotation measure estimated using average spectra
            }
\vspace{-3em}
\end{deluxetable}

\section{FRB Data and Analysis Methods} \label{sec:methods}

\qp\ \citep{dial25qp} and \newer\ \citep{Marnoch2023,twins} were detected in the Commensal Real-time ASKAP Fast Transients \citep[CRAFT;][]{Bannisteretal2017,shannon24craftics} survey on the Australian Square Kilometre Array Pathfinder \citep[ASKAP;][]{Hotan2021ASKAP}, using the Fast Real-time Engine for Dedispersing Amplitudes \citep[FREDDA;][]{2023MNRAS.523.5109Q}. The raw voltage data recorded upon detection were processed offline using the CRAFT Effortless Localization and Enhanced Burst Inspection pipeline \citep[CELEBI;][]{Scott2023_CELEBI}. Important observed and inferred properties of these two FRBs are listed in Table~\ref{tab:frbs}. Polarization calibration was done as part of post-processing, using the Vela pulsar (PSR J0835-4510) as the calibrator, for both FRBs. Details of the polarization analysis --- including construction of the dynamic spectra of Stokes' parameters I, Q, U, V and calculation of the linearly polarized flux density ($L$) --- are described in \citet{twins} and \citet{dial25qp}. The observed position angle (PA, $\chi_{\rm obs}$) of linear polarization, and the polarization ellipticity angle (EA, $\psi_{\rm obs}$), were calculated using the relations
\begin{align}
    \chi_{\rm obs} = \frac{1}{2} \tan^{-1}(U/Q) \\
    \psi_{\rm obs} = \frac{1}{2} \tan^{-1}(V/L) 
\label{eqn:polang}
\end{align}
where the PA is measured with respect to an arbitrary reference direction. We do not convert this to absolute position angle.

\subsection{Faraday Rotation} \label{sec:frot}

Faraday rotation measures (RM) of the FRBs were estimated following two different methods --- a linear fit to the variation of ${\rm PA_{obs}}$ with $\lambda^2$ and the technique of RM synthesis \citep{burn66rms, brentjens05rms} using the publicly available package RM Tools \citep{rmtools20} --- which yielded completely consistent results \citep[see][]{twins}. Results obtained from linear fits are used as the estimated values of RM. For all FRBs, the RM was measured from the time-averaged spectra over the entire burst, as well as over shorter time ranges to probe the RM variation across the bursts.

The observed $Q,U$ dynamic spectra were `corrected' (de-rotated) for the average RM for each FRB, applying the wavelength-dependent transformation 
\begin{equation}
	\begin{bmatrix}
		Q (\lambda, t) \\
		U (\lambda, t)
	\end{bmatrix}
	= 
	\begin{bmatrix}
		\cos{\xi (\lambda)} & -\sin{\xi (\lambda)} \\
		\sin{\xi (\lambda)} & \cos{\xi (\lambda)} 
	\end{bmatrix}    				
	\begin{bmatrix}
		Q (\lambda, t) \\
    		U (\lambda, t) 
	\end{bmatrix} _{\rm obs}	
	\label{eqn_defrot}
\end{equation}
where 
\begin{equation}
    \xi (\lambda) = -2 *{\rm RM_{avg}} * (\lambda^2 - \lambda_0^2)
\end{equation}
is the wavelength-dependent de-rotation angle and $\lambda_0$ is the reference wavelength, chosen to be the wavelength corresponding to the central frequency of the observing band.

\begin{figure*}[ht]
\centering
\includegraphics[scale=0.92,trim={0cm 0cm 0.15cm 0cm},clip]{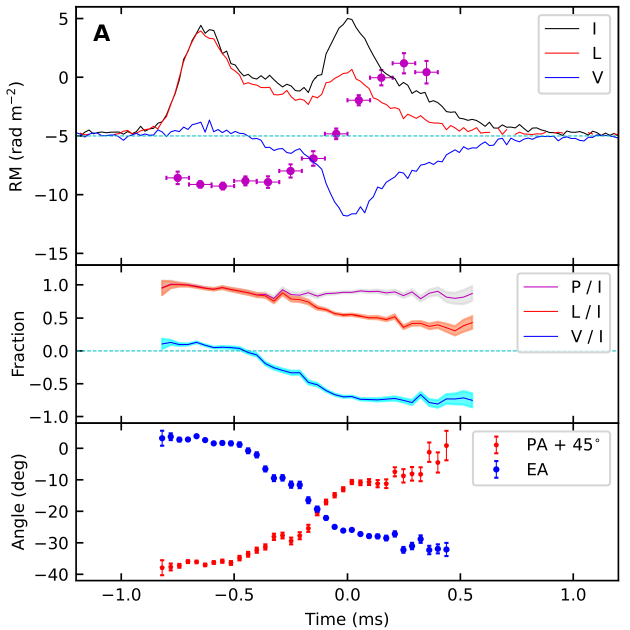}
\includegraphics[scale=0.92,trim={1.25cm 0cm 0cm 0cm},clip]{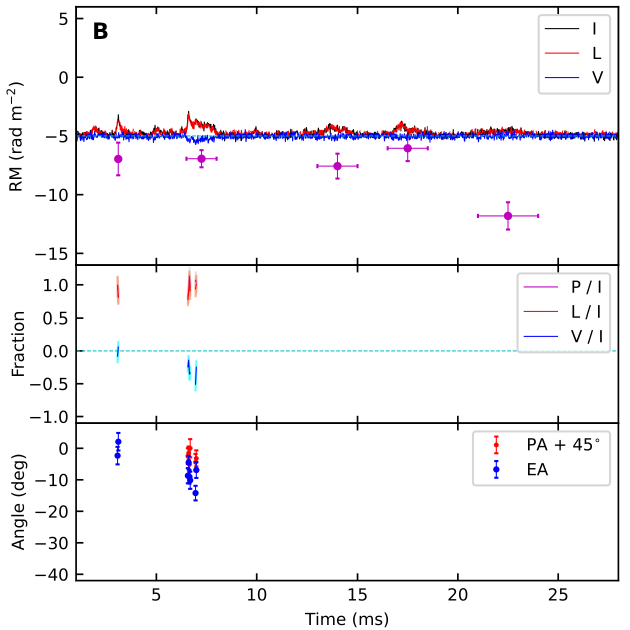}
\caption{\textbf{Frequency averaged polarization time profile of \qp\ in the primary sub-burst [A, \textit{left}] and in the fainter trailing sub-bursts [B, \textit{right}] after correcting for $\rm RM = -5.75\;rad\:m^2$.} The top panels show the normalized profiles of total intensity (I), linear polarization (L) and circular polarization (V), as well as estimates of RM in different time ranges (indicated by the x-errorbars). Note that the intensity axis has not been shown explicitly. The middle panels show profiles of fractional total (P), linear (L) and circular (V) polarization. The bottom panels show polarization position angle (PA, $\chi$) and ellipticity angle (EA, $\psi$). The time axis, with its reference (zero) set to the maximum of the intensity profile, has different scales in the left and the right panels.  
\label{fig_230708_fullpol575}}
\vspace*{1.0cm}
\end{figure*}

\subsection{Polarization Time Profiles} \label{sec:tprofile}

The de-rotated dynamic spectra of the Stokes parameters were averaged over frequency to generate the polarization time profiles at different time resolutions. Different time resolutions were chosen in different cases such that high S/N is retained in each time bin while the bins are small enough to probe the temporal evolution of polarization properties across the burst profiles. Estimates of the PA ($\chi$) and EA ($\psi$) were discarded where $L$ has a signal-to-noise ratio of S/N $ < 2$. 

We used the \pncr\ sphere representation \citep[e.g.][]{malykin97pspreview}, which is a unit sphere in the 3-dimensional polarization space with (Cartesian) co-ordinate axes defined by $Q/I$, $U/I$ and $V/I$, to investigate the temporal changes in the polarization state of the FRBs. Each point on the surface of the \pncr\ sphere represents a unique polarization state of fully polarized EM waves, with longitude $= 2 \chi$ and latitude $= 2 \psi$. Diametrically opposite points represent mutually orthogonal polarization states. Partially polarized EM waves can be represented by points beneath the surface, i.e.\ with radii smaller than unity. All FRBs discussed in this work are highly polarized with total polarization fractions $\gtrsim 70\%$ over their entire burst profiles. We hence neglect the ``depth'' (beneath the surface) and plot all points on the surface of the \pncr\ sphere using the measured values of $\chi$ and $\psi$.  

\begin{figure*}[htbp]
\centering
\includegraphics[scale=0.95,trim={0.0cm 0.0cm 0.0cm 0.0cm},clip]{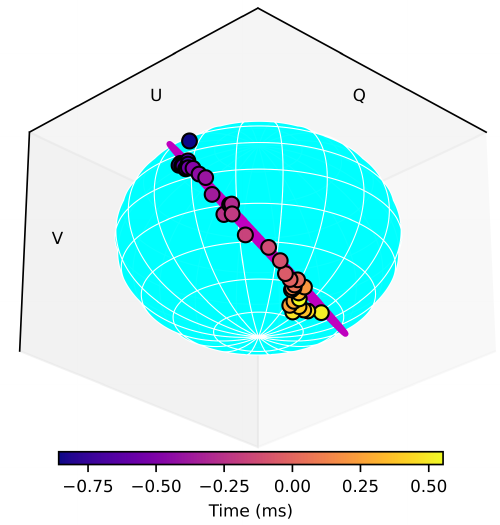}
\includegraphics[scale=0.95,trim={0.0cm 0.0cm 0.0cm 0.0cm},clip]{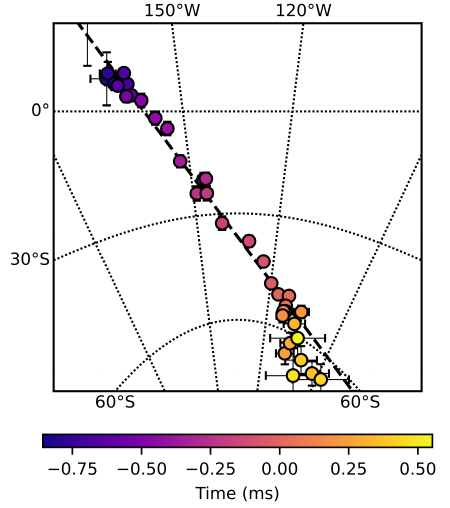}
\caption{\textbf{Temporal evolution of polarization state across the primary sub-burst of \qp\ on the \pncr\ sphere after correcting for $\rm RM = -5.75\;rad\:m^2$ shown in a 3-D plot (\textit{left}) and in gnomonic projection (\textit{right}).} Each data point represents the polarization state of emission in a 39 $\mu$s time bin. The best-fit great circle is shown as a solid magenta curve in the left panel and a dashed black curve in the right panel. Great circles on the surface of a sphere are transformed into straight lines in gnomonic projection.
\label{fig_230708_pspful_575}}
\end{figure*}

\section{\qp\ } \label{sec:qp}

\qp\ is a relatively bright event featuring a double-peaked primary sub-burst followed by trailing faint quasi-periodic sub-bursts, as shown in Figure~\ref{fig_230708_fullpol575}. Dynamic spectra and other details of this FRB have been presented in \citet{dial25qp}. The primary sub-burst has an average rotation measure of $\rm RM_{avg} = -5.75 \pm 0.20 \;rad\:m^2$, which is consistent with the RM estimates for the trailing sub-bursts\footnote{The same data that has been used by \citet{dial25qp} were independently analyzed for this work. The results are consistent with those presented in \citet{dial25qp}. The difference in the reported values of the average RM resulted from the choice of slightly different time windows for RM estimation and the ``apparent'' temporal variation of RM.}. However, polarization spectra integrated over shorter time ranges show a smooth temporal variation of apparent RM across the primary sub-burst.

We de-rotated the $Q - U$ dynamic spectra of \qp\ for $\rm RM_{avg}$ of the primary sub-burst. The frequency-averaged time profile shows a very high degree of polarization. The primary sub-burst exhibits time dependent gradual conversion between linear and circular polarization --- with the first peak almost fully linearly polarized and the second peak showing comparable fractions of linear and circular polarization (see the middle pannel in Figure~\ref{fig_230708_fullpol575}). The PA of linear polarization ($\chi$) and the EA ($\psi$) show smooth correlated temporal evolution across the primary sub-burst of \qp, indicating a gradual evolution of the polarization state of radio emission.

\begin{figure*}[h]
\centering
\includegraphics[scale=0.71,trim={0.0cm 0cm 0.0cm 0.0cm},clip]{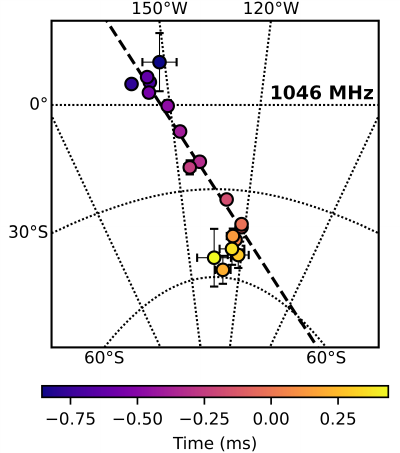}
\includegraphics[scale=0.71,trim={0.0cm 0cm 0.0cm 0.0cm},clip]{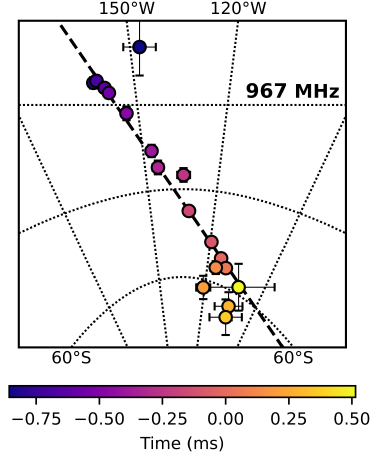}
\includegraphics[scale=0.71,trim={0.0cm 0cm 0.1cm 0.0cm},clip]{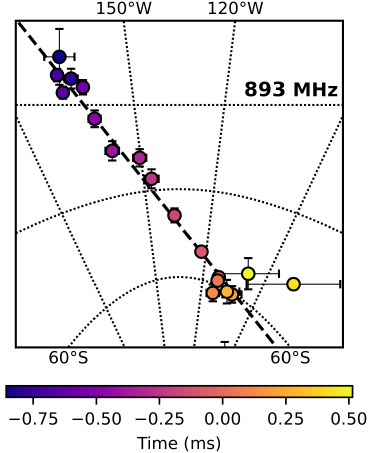}
\includegraphics[scale=0.71,trim={0.1cm 0cm 0.1cm 0.0cm},clip]{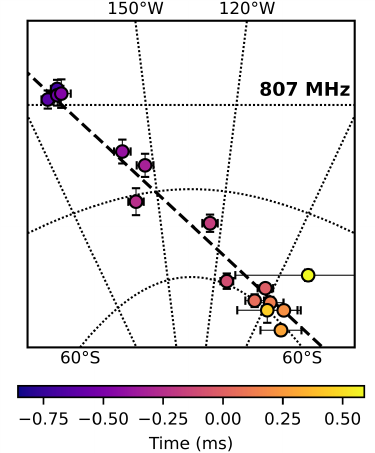}
\includegraphics[scale=0.71,trim={0.0cm 0cm 0.0cm 0.0cm},clip]{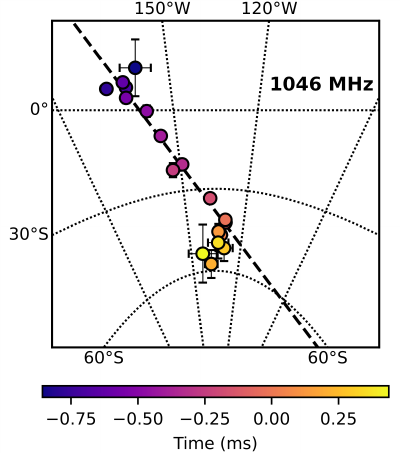}
\includegraphics[scale=0.71,trim={0.0cm 0cm 0.0cm 0.0cm},clip]{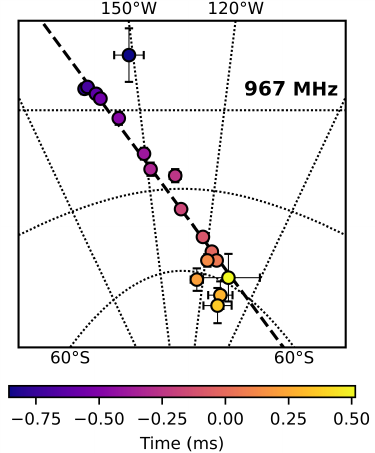}
\includegraphics[scale=0.71,trim={0.0cm 0cm 0.1cm 0.0cm},clip]{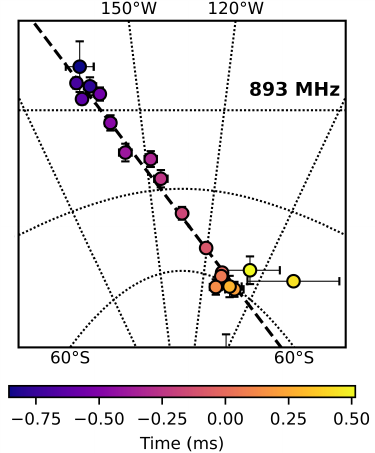}
\includegraphics[scale=0.71,trim={0.1cm 0cm 0.1cm 0.0cm},clip]{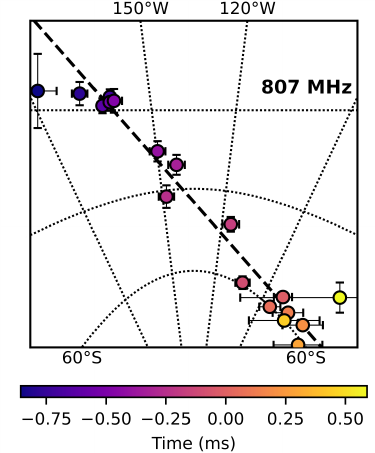}
\caption{\textbf{Temporal evolution of polarization state across the primary sub-burst of \qp\ on the \pncr\ sphere after correcting for $\rm RM = -5.75\;rad\:m^2$ [\textit{top} panels] and $\rm RM = -8.85\;rad\:m^2$ [\textit{bottom} panels] in different frequency sub-bands centred at (from left to right) 1045.5 MHz, 966.6 MHz, 893.2 MHz and 806.6 MHz, respectively, shown in gnomonic projection.} Each data point represents the polarization state of radio emission averaged over a 78 $\mu$s time bin. The best-fit great circle is shown as a dashed black curve in each panel.   
\label{fig_230708_pspsb}}
\end{figure*}

\begin{figure*}[ht]
\centering
\includegraphics[width=0.49\linewidth]{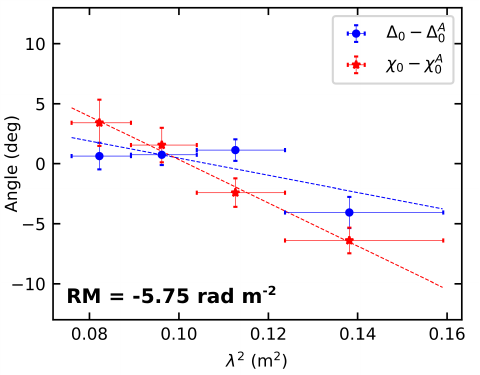}
\includegraphics[width=0.49\linewidth]{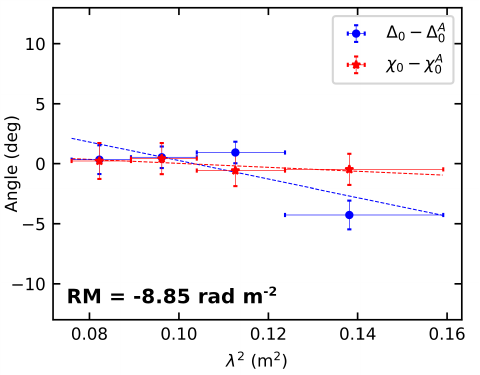}
\caption{\textbf{Wavelength dependence of the best-fit great circle parameters ($\Delta_0$, $\chi_0$) in \qp\ after correcting for $\rm RM = -5.75\;rad\:m^2$ [A, \textit{left}] and $\rm RM = -8.85\;rad\:m^2$ [B, \textit{right}], respectively.} $\Delta_0^A$ and $\chi_0^A$ are the best-fit parameters for the full-band, i.e.\ $\Delta_0^A = 61.4\dg$, $\chi_0^A=10.2\dg$ for the left panel and $\Delta_0^A = 61.6\dg$, $\chi_0^A=9.4\dg$. The dashed lines show best-fit linear variations of the corresponding angles with $\lambda^2$.  
\label{fig_230708_pspec}}
\end{figure*}

\subsection{Trajectory on the \pncr\ Sphere} \label{sec:gcfit}

To investigate the temporal evolution of the polarization state of the primary sub-burst of \qp\ we time-averaged the data over bins of 39 $\mu$s. The trajectory of the polarization state on the surface of the \pncr\ sphere, plotted following the method described in Section~\ref{sec:tprofile}, is found to be well-described by a great circle, as shown in Figure~\ref{fig_230708_pspful_575}. A general great circle on the surface of the \pncr\ sphere is described by
\begin{equation}
\tan{2\psi} = a \cos{2\chi} + b \sin{2\chi},
    \label{eqn_mc2}
\end{equation}
which is the intersection of the \pncr\ sphere and a plane passing through its centre, with the parameters $a$ and $b$ determining the orientation of the plane in space. Defining 
\begin{equation*}
    \Delta_0 = \tan^{-1} \left( \sqrt{a^2+b^2} \right) \; {\rm and} \; \chi_0 = -\frac{1}{2} \tan^{-1} \left( {\frac{a}{b}} \right),
\end{equation*}
a general great circle can be parameterized as
\begin{equation}
\tan(2 \psi) = \tan{\Delta_0} \; \sin{(2\chi - 2\chi_0)}
    \label{eqn_mc}
\end{equation}
where $\Delta_0$ is the inclination of the great circle with respect to the equatorial plane, and $\chi_0$ is a reference value of PA such that $\chi = \pm \chi_0$ gives the intersections between the great circle and the equator ($Q-U$ plane). The motion along the great circle can be parameterized by 
\begin{equation}
    \Gamma(t) = \frac{1}{2} \arctan \left( \frac{\sqrt{U'^2 + V^2}}{-Q'} \right)
    \label{eqn_gamapar}
\end{equation}
with
\begin{equation}
     \begin{bmatrix}
		Q' \\
		U'
	\end{bmatrix}
	= 
	\begin{bmatrix}
		\cos{2\chi_0} & \sin{2\chi_0} \\
		-\sin{2\chi_0} & \cos{2\chi_0} 
	\end{bmatrix}    				
	\begin{bmatrix}
		Q  \\
    	U 
	\end{bmatrix} 
    \label{eqn_qurot}
\end{equation}
where $2\Gamma$ represents the angular distance from the equator along the specific great circle defined by $\Delta_0$ and $\chi_0$.

\begin{figure*}[t!]
\centering
\includegraphics[scale=0.92,trim={0cm 0cm 0.15cm 0cm},clip]{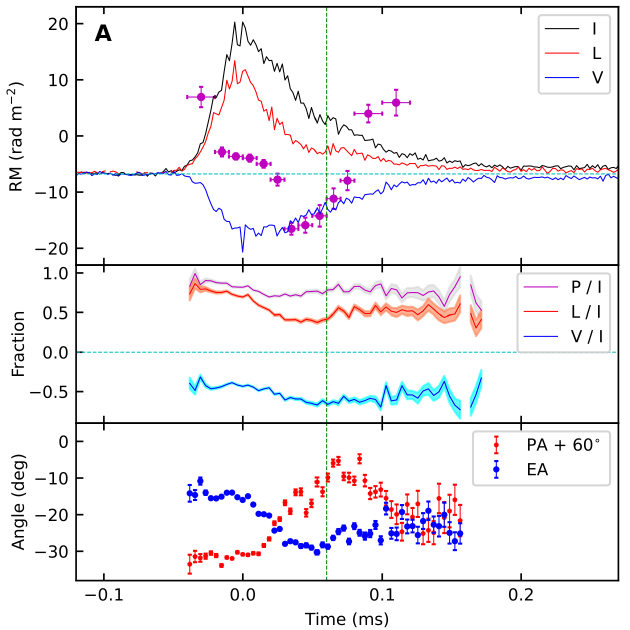}
\includegraphics[scale=0.92,trim={1.25cm 0cm 0cm 0cm},clip]{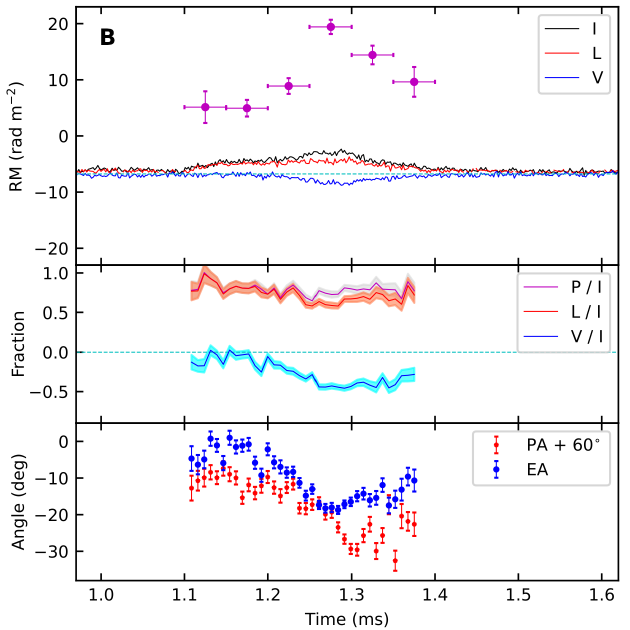}
\caption{\textbf{Frequency averaged polarization time profile of \newer\ in sub-burst $A$ [A, \textit{left}] and sub-burst $B$ [B, \textit{right}] after correcting for $\rm RM = 4.55\;rad\:m^2$.} Note that the intensity axis has not been explicitly shown in the top panels. The time axis has different scales in the left and the right panels. The vertical dashed line at $t=0.06$ ms in the left panel separates the `peak' and the `tail' of the sub-burst $A$, as discussed in Section~\ref{sec:newer}. See Figure~\ref{fig_230708_fullpol575} for a description of the symbols.
\vspace*{0.2cm}
\label{fig_210912_fullpol455}}
\end{figure*}

The best-fit\footnote{The best-fit great circle parameters were estimated using Equation~\ref{eqn_mc2} and the method of least-squares.} great circle to the trajectory of \qp\ has an inclination of $\Delta_0 = 61.4 \pm 0.5 \dg$ and a reference PA of $\chi_0 = 10.2 \pm 0.7 \dg$. The polarization vector rotates by $\sim 90\dg$ along the great circle across the primary sub-burst of \qp. At the ``tail'' of the primary sub-burst, the trajectory of the polarization vector slightly deviates from the best-fit great circle. The fainter trailing sub-bursts do not have sufficient S/N for a similar analysis.

\subsection{Frequency Dependence of the Trajectories} \label{sec:freqdep}

The temporal variation of apparent RM across the primary sub-burst implies that residual frequency dependence remains in the $Q, U$ dynamic spectra even after de-rotating for the average RM. The high detection S/N and the high polarization fraction of \qp\ allows us to investigate the frequency dependence of the great circle trajectories, by dividing the observing frequency band in multiple sub-bands and performing similar analysis for each sub-band. We divided the observing frequency band in four (unequal) sub-bands centred at 1045.5 MHz, 966.6 MHz, 893.2 MHz and 806.6 MHz, respectively, and calculated sub-band time profiles by frequency-averaging over each sub-band at a time resolution of 78 $\mu$s. The band widths of the sub-bands were chosen such that each sub-band has sufficient S/N. 

The trajectory of the polarization vector in each of the sub-bands across the primary sub-burst is well described by a great circle on the \pncr\ sphere, as shown in the top panels of Figure~\ref{fig_230708_pspsb}, although a small deviation is observed at the fading tail (similar to the observed deviations for the full band profile). We estimated the best-fit great circle parameters ($\Delta_0$, $\chi_0$) for each sub-band and fit their frequency dependence with a linear function of $\lambda^2$, as shown in Figure~\ref{fig_230708_pspec}A. We find that the inclination only shows a marginally significant evidence of frequency dependence with a best fit relation $\rm \Delta_0 \sim - (1.2 \pm 0.5) \lambda^2$, while the reference PA varies with wavelength as $\rm \chi_0 \sim - (3.1 \pm 0.6) \lambda^2$, where the slopes of the relations have units of rad m$^{-2}$.

\section{\newer\ } \label{sec:newer}

\begin{figure*}[ht]
\centering
\includegraphics[scale=1,trim={0cm 0.1cm 0cm 0cm},clip]{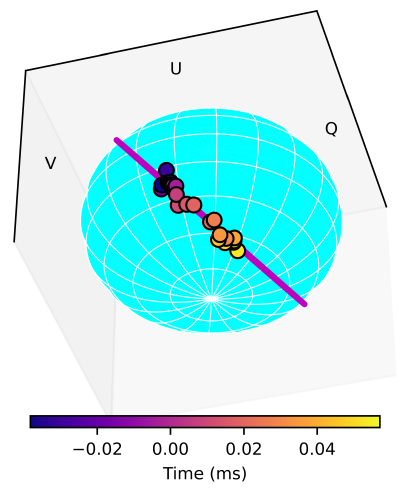}
\includegraphics[scale=1,trim={0cm 0cm 0cm 0cm},clip]{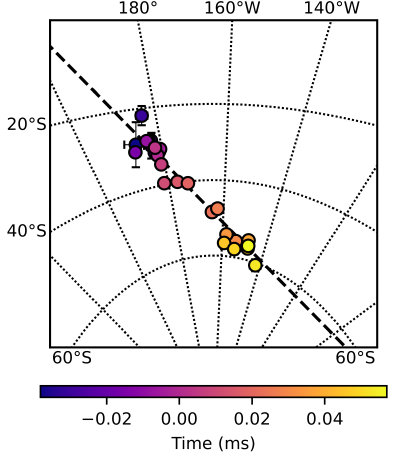}
\caption{\textbf{Temporal evolution of the polarization state across the `peak' of sub-burst $A$ of \newer\ on the \pncr\ sphere after correcting for $\rm RM = 4.55\;rad\:m^2$ shown in a 3-D plot (\textit{left}) and in gnomonic projection (\textit{right}).} Each data point represents the polarization state in a 3.8 $\mu$s time bin. The best-fit great circle is shown as a solid magenta curve in the left panel and a dashed black line in the right panel.
\vspace*{0.5cm}
\label{fig_210912_apk_pspful_455}}
\end{figure*}

\newer\ is another bright FRB featuring a strong primary sub-burst ($A$) followed by a relatively weaker secondary sub-burst ($B$), as shown in Figure~\ref{fig_210912_fullpol455}. We refer to \citet{twins} for dynamic spectra and other details of this FRB. The average RM of \newer, measured over the entire burst profile, is $\rm RM_{avg} = 4.55 \pm 0.49 \;rad\:m^2$. However, as discussed in \citet{twins}, the apparent RM varies across the two sub-bursts (see the top panels in Figure~\ref{fig_210912_fullpol455}). We de-rotated the $Q - U$ dynamic spectra of \newer\ for $\rm RM_{avg} = 4.55\;rad\:m^2$ following the method described in Section~\ref{sec:frot}. The frequency-averaged time profile shows high degree of polarization and both sub-bursts exhibit time-dependent gradual conversion between linear and circular polarization (middle panels in Figure~\ref{fig_210912_fullpol455}). The PA ($\chi$) and the EA ($\psi$) show smooth correlated temporal evolution across both sub-bursts, although the nature of correlation is different in the two sub-bursts (bottom panels in Figure~\ref{fig_210912_fullpol455}). 

As noted by \citet{twins}, the fading `tail' of sub-burst $A$ consists of a trailing component with different spectral and polarization properties compared to the `peak' of the sub-burst. The PA and EA trends can be seen to change after the trailing component becomes dominant in the time profile. We hence investigate the polarization properties of the `peak' and `tail' of sub-burst $A$ separately, using a boundary at $t=0.06$ ms (with respect to the maximum of the total intensity profile) that roughly separates the trailing component from the brightest component. We note that sub-burst $B$ also exhibits changes in PA and EA trends at the tail; however, its relative faintness does not allow us to identify any clear component separation\footnote{We note that in case of \qp, the slight deviation of the polarization state from the great circle trajectory at the fading tail of the primary sub-burst --- shown in Fig.~\ref{fig_230708_pspful_575} --- could also be caused by the presence of a faint emission component with different polarization properties.}.

\subsection{Polarization State in the Two Sub-bursts} \label{sec:polab}

We used a time resolution of 3.8 $\mu$s to probe the temporal evolution of the polarization state in the peak of sub-burst $A$ of \newer. Like the primary sub-burst of \qp, the frequency-averaged trajectory of the polarization state of the peak of sub-burst $A$, plotted on the surface of the \pncr\ sphere following the method described in Section~\ref{sec:tprofile}, is found to be well described by a great circle, as shown in Figure~\ref{fig_210912_apk_pspful_455}. The best-fit great circle has an inclination of $\Delta_0^A = 62.4 \pm 1.4 \dg$ and a reference PA of $\chi_0^A = -10.4 \pm 1.3 \dg$. 

\begin{figure*}[ht]
\centering
\includegraphics[scale=1,trim={0.0cm 0cm 0.0cm 0cm},clip]{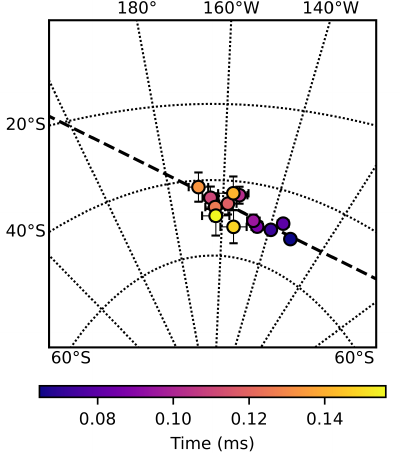}
\includegraphics[scale=1,trim={0.0cm 0cm 0.0cm 0cm},clip]{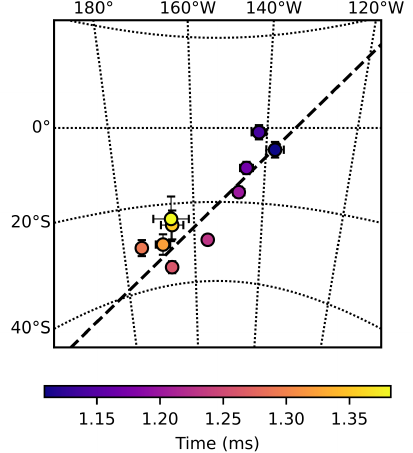}
\caption{\textbf{Temporal evolution of the polarization state across the `tail' of sub-burst $A$ [A, \textit{left}] and across sub-burst $B$ [B, \textit{right}] of \newer\ on the \pncr\ sphere after correcting for $\rm RM = 4.55\;rad\:m^2$.} Each data point represents the polarization state of emission in a 7.6 $\mu$s time bin in the \textit{left} panel and 46 $\mu$s time bin in the \textit{right} panel. The best-fit great circle is shown as a dashed black line in each panel.
\label{fig_210912_atlb_pspful_455}}
\end{figure*}

Owing to the relative faintness of the tail of sub-burst $A$, we used a coarser time resolution of 7.6 $\mu$s to probe its polarization state which shows a relatively weaker evolution compared to the peak of the sub-burst. The polarization vector moves in opposite direction on the \pncr\ sphere compared to the peak along a slightly different track, as shown in Figure~\ref{fig_210912_atlb_pspful_455}A. The best-fit great circle to the trajectory of the tail is given by $\Delta_0 = 50.9 \pm 1.9 \dg$ and $\chi_0 = -19.4 \pm 2.2 \dg$. 

As shown in Figure~\ref{fig_210912_atlb_pspful_455}B, the nature of the evolution of the polarization state in sub-burst $B$ of \newer\ is different from that in sub-burst $A$. The frequency-averaged trajectory of the polarization vector, plotted on the surface of the \pncr\ sphere at a time resolution of 30 $\mu$s in Figure~\ref{fig_210912_atlb_pspful_455}B, is completely different from that of sub-burst $A$. The best-fit great circle has $\Delta_0^B = 47.8 \pm 3.4 \dg$, $\chi_0^B = -66.5 \pm 3.9 \dg$, and a different orientation on the \pncr\ sphere than that corresponding to sub-burst $A$.   

\subsection{Frequency Dependence of Trajectories} \label{sec:freqdepnewer}

\begin{figure*}[ht]
\centering
\includegraphics[scale=0.74,trim={0.0cm 0cm 0.2cm 0cm},clip]{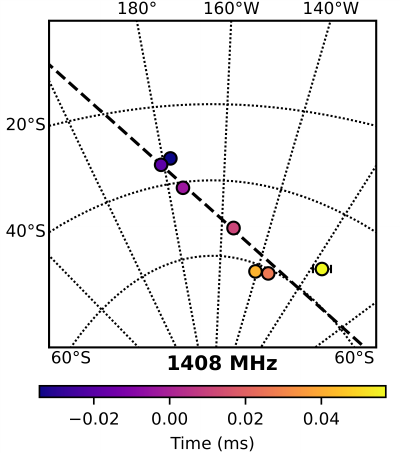}
\includegraphics[scale=0.74,trim={0.4cm 0cm 0.4cm 0cm},clip]{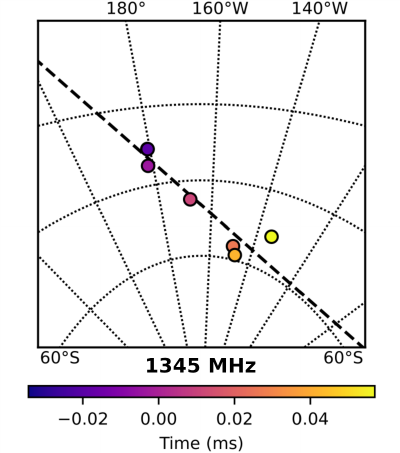}
\includegraphics[scale=0.74,trim={0.5cm 0cm 0.3cm 0cm},clip]{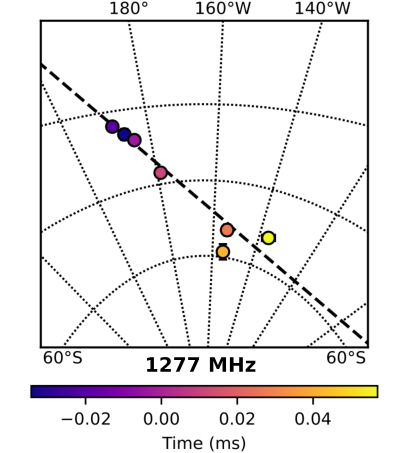}
\includegraphics[scale=0.74,trim={0.5cm 0cm 0.3cm 0cm},clip]{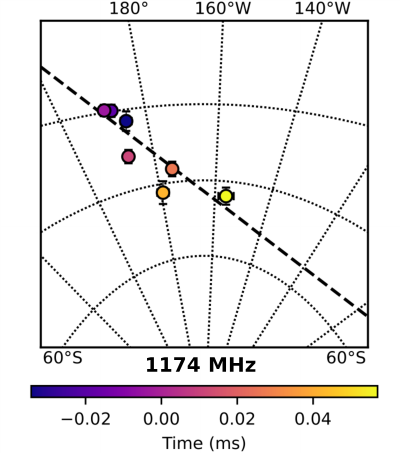}
\caption{\textbf{Temporal evolution of polarization state across the `peak' of sub-burst $A$ of \newer\ on the \pncr\ sphere after correcting for $\rm RM = 4.55\;rad\:m^2$ in different frequency sub-bands centred at (from left to right) 1408.0 MHz, 1345.0 MHz, 1276.8 MHz and 1174.4 MHz, respectively.} Each data point represents the polarization state of radio emission averaged over a 15 $\mu$s time bin. The best-fit great circle is shown as a dashed black line in each panel.   
\vspace*{0.75cm}
\label{fig_210912_apk_pspsb_455}}
\end{figure*}

The high S/N of the peak of sub-burst $A$ allows us to probe the frequency dependence of the evolution of the polarization state, following the same method used for \qp. We divided the observing frequency band in four unequal sub-bands centred at 1408.0 MHz, 1345.0 MHz, 1276.8 MHz and 1174.4 MHz, respectively, and calculated sub-band time profiles by frequency-averaging over each sub-band at a time resolution of 15 $\mu$s. The tail of sub-burst $A$ and sub-burst $B$ have much lower S/N compared to the peak of sub-burst $A$. We hence divided the observing frequency band in two unequal sub-bands centred at 1339.8 MHz and 1174.4 MHz, respectively, and calculated sub-band time profiles by frequency-averaging over each sub-band at a time resolution of 15 $\mu$s for the tail of sub-burst $A$ and 46 $\mu$s for sub-burst $B$. The trajectories of the polarization vector in each sub-band and the corresponding best-fit great circles are shown in Figure~\ref{fig_210912_apk_pspsb_455} for the peak of sub-burst $A$, and in Figures~\ref{fig_210912_atlb_pspsb_455} for the tail of sub-burst $A$ and for sub-burst $B$.

\begin{figure}[t]
\centering
\includegraphics[width=0.98\linewidth]{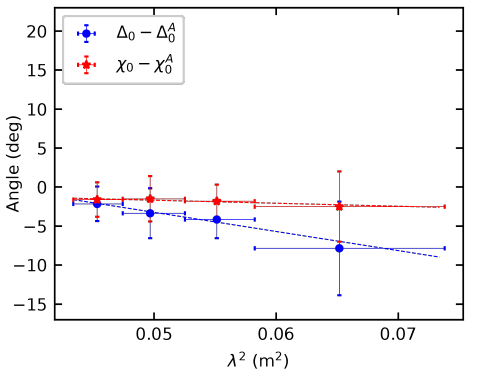}
\caption{\textbf{Wavelength dependence of the best-fit great circle parameters ($\Delta_0$, $\chi_0$) in sub-burst $A$ of \newer\ after correcting for $\rm RM = 4.55\;rad\:m^2$.} Best-fit parameters for the full-band have been subtracted from these measurements. The dashed lines show the best linear fits to the corresponding angles with $\lambda^2$.  
\label{fig_210912_pspeca}}
\end{figure}

The inclination ($\Delta_0$) and the reference PA ($\chi_0$) of the best-fit great circles for the peak of sub-burst $A$ show no significant frequency dependence across the observing band, as seen in Figure~\ref{fig_210912_pspeca}, with $\rm \Delta_0 \sim - (4.3 \pm 4.5) \lambda^2$, while the reference PA varies with wavelength as $\rm \chi_0 \sim - (0.7 \pm 3.9) \lambda^2$, where the slopes have units of rad m$^{-2}$. The frequency dependence of these parameters for the tail of sub-burst $A$ could not be constrained from the data. Sub-burst $B$ of \newer\ has $\rm \Delta_0 \sim - (21.7 \pm 6.0) \lambda^2$ and $\rm \chi_0 \sim - (7.4 \pm 10.1) \lambda^2$, showing a weak frequency dependence of $\Delta_0$ (see Figure~\ref{fig_210912_pspecb}).

\section{Possible Interpretations} \label{sec:model}

The observed trajectories of the polarization state in \qp\ and \newer\ --- which results from correlated variations of linear and circular polarization --- are unlikely to be a random coincidence, although a formal estimation of the probability is not straightforward and critically depends on the assumptions made. Instrumental effects also appear unlikely to lead to the observed trajectories of the polarization vector as discussed in detail in Section~\ref{sec:instrument}. 

The origin of the circular polarization in FRBs --- whether it is an intrinsic property of radio emission or an effect of propagation through magneto-ionic media \citep[e.g.][]{mckinnon24nopm,kennet98gfr,melrose04cp,lyutikov22apj,macquart00cir,macquart00circ} --- remains unclear. Continuous variation of the polarization state across the burst, with changing fractions of linear and circular polarization and the polarization vector tracing smooth curves on the \pncr\ sphere, has been previously observed for other FRBs \citep{Sherman2023,cho20apjl}. \citet{mandlik24ml} reported observation of \utfrb\ with the  UTMOST-NS facility at the Molonglo radio telescope, which shows the polarization vector tracing a segment of a great circle on the \pncr\ sphere as seen in \qp\ and \newer. Circular polarization induced by scintillation in a turbulent magnetized medium \citep{macquart00circ} is unlikely to lead to such smooth trajectories on the \pncr\ sphere. `Generalized' Faraday rotation (or Faraday conversion) has been suggested as a possible origin of varying degree of circular polarization \citep{cho20apjl,kumar22gfr,uttarkar24cp}. However, as discussed by \citet{dial25qp}, the time-averaged polarization spectra of \qp\ do not show any conclusive evidence of `generalized' Faraday rotation \citep{lower21gfr,uttarkar24cp}. Here we discuss three possible scenarios that may lead to the observed great circle trajectories of the polarization vector on the \pncr\ sphere for \qp\ and \newer\ --- transition between two orthogonal polarization modes and the effects of magneto-ionic birefringence.

\subsection{Polarization mode transition} \label{sec:onpm}

The great circle trajectories observed in the FRBs could be explained by a transition between the natural modes of wave propagation in a magneto-ionic plasma \citep{dyks21coh, mckinnon24nopm}. The transition is caused by a change in the relative strength of the modes' electric fields. The trajectory of a mode transition on the \pncr\ sphere follows the geodesic that connects the orientations of the mode polarization vectors \citep{mckinnon24nopm}. Therefore, the maximum angular extent of a transition on the sphere is less than or equal to $180\dg$. Assuming the observed trajectories are caused by mode transitions, their limited angular extent suggests the transitions are incomplete. The large and nearly constant polarization fraction of the FRB emission indicates the polarization modes are highly, but not completely, coherent. If the modes were completely coherent, the polarization fraction would remain constant at unity over the duration of the burst \citep{dyks21coh}. If the coherence of the modes was low, the polarization fraction would vary across the burst with the EA reaching a maximum when the polarization fraction is minimum \citep{oswald23pcoh, mckinnon24nopm}. The polarization minimum occurs when the electric field strengths of the modes are equal. The observations do not show clear associations between polarization minima and EA maxima. While the absence of this association could be attributed to an incomplete transition, the combination of the nearly constant polarization fraction and large EA variations suggests the great circle trajectories are not likely to be caused by polarization mode transitions.

\subsection{Effects of Elliptical Birefringence} \label{sec:ebm}

FRB progenitors are thought to be surrounded by a magneto-ionic shell/nebula, such as a supernova remnant or pulsar wind nebula \citep[e.g.][]{Chatterjee2017,Michilli2018_121102,bruni24prs}. \citet{vedantham19gfr} and \citet{gruzinov19gfr} have suggested that linearly polarized FRB emission can be (partially) converted to circular polarization as the burst propagates through the nebula. These polarization models generally assume that the plasma in the nebula is comprised of a mixture of thermal and relativistic particles, such that the plasma's natural modes of wave propagation are elliptically polarized. The ellipticity of the natural modes is determined by the ratio of the number densities of the thermal and relativistic particles in the plasma \citep{kennet98gfr}. The ellipticity is $\epsilon=\tan(\psi_o)$, where $\psi_o$ is the ellipticity angle of the mode polarization vectors. If the modes are linearly polarized, $\psi_o = 0$, $\epsilon=0$, and the plasma particles are relativistic. If the modes are circularly polarized, $\psi_o = \pm\pi/4$, $\epsilon=\pm 1$, and the plasma particles are thermal, as in the interstellar medium. An elliptically birefringent medium alters the phase offset, $\eta$, between its elliptically polarized natural modes of propagation and generally causes the trajectory of the observed polarization vector to trace a small circle on the Poincar\'e sphere \citep[see, e.g., Figure 3 of][]{kennet98gfr}. The mode polarization vectors form the normal to the plane defined by the small or great circle, and the motion along the circle is associated with temporal variation of the phase offset, $\eta$. The relationship between $\psi_o$ and the observed inclination angles of the great circles (GCs), $\Delta$, is therefore $\psi_o = (\pi/2-\Delta)/2$. The ellipticity determined from the inclination angles observed in \qp\ and the primary sub-burst of \newer\ is $\epsilon \approx 0.25$, indicating the plasma particles are primarily relativistic.

\begin{figure}[t]
\centering
\includegraphics[width=0.98\linewidth]{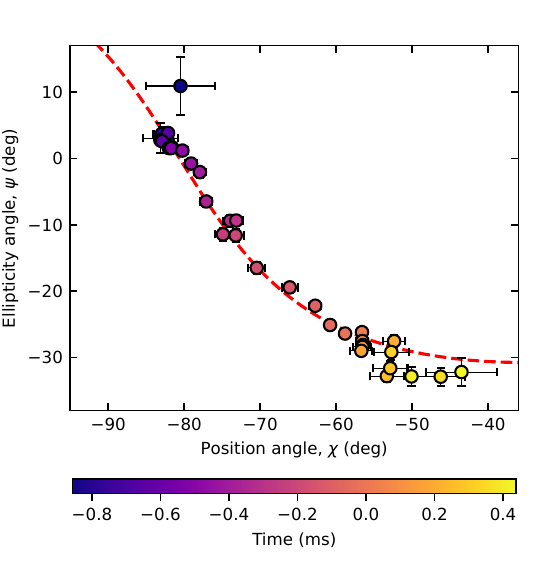}
\caption{\textbf{Comparison of the measured PAs and EAs in \qp\ with those predicted by the elliptical birefringence model.} The measured values of the angles are shown by the filled circles, and the model PAs and EAs calculated from Equations~\ref{eqn:PA} and~\ref{eqn:EA} is shown by the red dashed line. The mode phase offset, $\eta$, in the model varies from $-20^\circ$ to $70^\circ$ in the figure.
\vspace*{0.8cm}
\label{fig_paea_gfr}}
\end{figure}

The observed GCs constrain the geometry of the mode polarization vectors relative to the polarization vector of the radiation incident upon the plasma. For the vector to trace a GC, the polarization vector of the incident radiation must be perpendicular to the mode polarization vectors. If the incident radiation is linearly polarized, its polarization vector is further constrained to reside along the line formed by the intersection of the GC with the equatorial plane of the sphere.

The variations in the PA and EA with $\eta$ due to elliptical birefringence are given by Equations~\ref{eqn:PA} and~\ref{eqn:EA} in Appendix~\ref{app:paeagfr}. Figure~\ref{fig_paea_gfr} compares the measured values of the PA and EA for \qp\ with those calculated from the equations. The figure shows that the elliptical-birefringence model can reproduce the observed GC trajectories. However, a potential problem with this interpretation of the observed GC trajectories is that the polarization of successive sub-bursts should traverse the same GC, assuming the physical properties of the nebula remain constant over $\sim$ms timescales, while the observed GCs for successive sub-bursts of \newer\ are different. Besides, the specific requirement of the polarization vector of the incident radiation to be perpendicular (on the \pncr\ sphere) to the polarization modes of the plasma in a surrounding nebula gives rise to a `fine-tuning' problem. Finally, the motion along the GC trajectories requires changes in the phase offset ($\eta$) between the two polarization modes of the plasma in sub-millisecond time scales keeping the polarization modes unchanged, which appear implausible in a realistic scenario. Therefore, propagation through an elliptically birefringent shell/nebula, comprised of a mixture of thermal and relativistic particles, appears unlikely to be the cause of the observed GC trajectories for \qp\ and \newer.  

\begin{figure*}[t]
\centering
\includegraphics[width=0.49\linewidth]{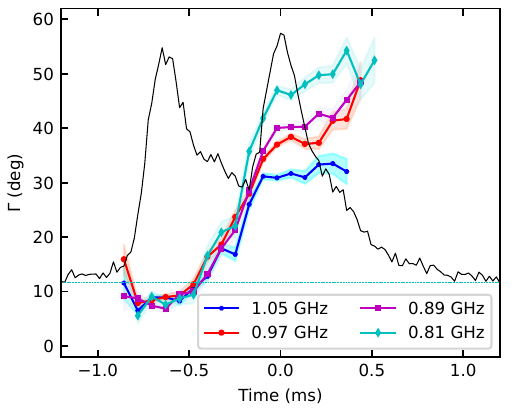}
\includegraphics[width=0.49\linewidth]{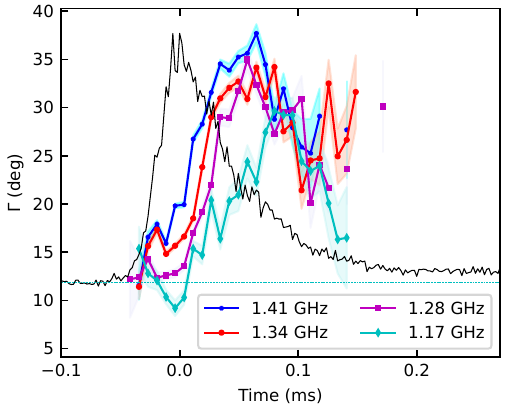} 
\caption{\textbf{Temporal variations of $\Gamma$ across the primary sub-bursts of \qp\ [A, \textit{left}] and \newer\ [B, \textit{right}] in four frequency sub-bands.} The central frequencies of the sub-bands are mentioned in the legend. Corrections for $\rm RM = -8.85\;rad\:m^2$, and  $\rm RM = 4.55\;rad\:m^2$ have been applied to the left and the right panels, respectively. Each data point represents the polarization state of radio emission averaged over a 78 $\mu$s time bin in the left panel and over a 7.6 $\mu$s time bin in the right panel. 
\vspace*{0.4cm}
\label{fig_bothgama}}
\end{figure*}
   
\subsection{Effects of Linear Birefringence} \label{sec:lbm}

Propagation through a linear birefringent medium (LBM), which introduces a phase difference between its two linearly polarized orthogonal propagation modes, may also induce significant circular polarization in (originally) linearly polarized emission. We assume that the incident radiation is fully linearly polarized with a position angle of $\gamma (t)$ -- i.e.\ $\gamma$ is the angle between the electric field on the sky plane and an arbitrarily chosen reference axis. The intrinsic time dependence of $\gamma$ can be attributed to the properties of the emission source (see Section~\ref{sec:gama}). An LBM introduces a phase difference $\delta$ between the two orthogonal linear components of the electric field along its optical axes (i.e.\ the `fast' axis and the `slow' axis). The angle between the reference axis and the fast axis of the LBM is denoted by $\phi$. Here we consider a scenario where the LBM is close to the FRB emission site, while the interstellar medium (ISM) and the intergalactic medium (IGM) only introduce Faraday rotation of the PA of the linear polarization by a frequency dependent angle, $\alpha (\lambda)$. The observed polarization vector, expressed in terms of $\chi$ (PA) and $\psi$ (EA), follows the relation (see Appendix~\ref{app:gcmod})
\begin{equation}
    \tan{2\psi} = \tan{\delta} \: \sin{(2\chi - 2\alpha + 2\phi)}. \label{eq:lbmgc}
\end{equation}
which represents a great circle on the surface of the \pncr\ sphere (Equation~\ref{eqn_mc}) with $(\delta, \alpha - \phi)$ $= (\Delta_0 +n\pi, \chi_0$) or $(n\pi - \Delta_0, \chi_0 \pm \pi/2)$ for arbitrary integral values of $n$.

\section{Great Circles from Linear Birefringence} \label{sec:gcfreq}

Assuming that the great circle trajectories of \qp\ and \newer\ resulted from propagation through an LBM, the inclination ($\Delta_0$) of the great circle with respect to the equator represents the phase difference ($\delta$) between the two orthogonal linear modes of the LBM, and is expected to depend on the frequency of the EM waves \cite[e.g.][]{lyutikov22apj}. As described in Section~\ref{sec:freqdep} and Figure~\ref{fig_230708_pspec}A, the frequency dependence of $\Delta_0$ could not be measured from our current observations and the limits can be used to put constraints on the magneto-ionic properties of the LBM (see Section~\ref{sec:lbmorigin}).

However, the reference PA ($\chi_0$) shows measurable frequency dependence for \qp, as described in Figure~\ref{fig_230708_pspec}A and Section~\ref{sec:freqdep}, which can be attributed to the frequency dependence of $\alpha$ as the orientation of the optical axes ($\phi$) of the LBM is expected to be frequency independent. We note that the $Q-U$ dynamic spectra have been de-rotated for the $\rm RM_{avg}$. However, as the RM of these FRBs appear to vary within the burst, $\rm RM_{avg}$ may not represent the ``true" degree of Faraday rotation introduced by the ISM/IGM, and the measured frequency dependence of $\chi_0$ would reflect the residual RM after correcting for $\rm RM_{avg}$. For \qp, the measured frequency dependence of $\chi_0$ suggests a residual RM of $-3.1 \pm 0.6$ rad m$^{-2}$, which in turn implies an inferred true RM $= -8.85 \pm 0.63$ rad m$^{-2}$. This estimate of the true RM is consistent with the observed RM of the first peak  of the primary sub-burst (see Figure~\ref{fig_230708_fullpol575}A) which exhibits a very high degree of linear polarization and a small fraction of circular polarization.

We repeated the entire analysis described in Section~\ref{sec:qp} after independently de-rotating the observed $Q-U$ dynamic spectra with $\rm RM = -8.85$ rad~m$^2$ using Equation~\ref{eqn_defrot}. The frequency-averaged polarization time profiles (see Figures~\ref{fig_230708_fullpol885}) show only small differences from the results presented in Figures~\ref{fig_230708_fullpol575} and \ref{fig_230708_pspful_575}. As shown in the bottom panels of Figure~\ref{fig_230708_pspsb} and Figure~\ref{fig_230708_pspec}B, the \pncr\ sphere trajectory --- specifically the reference PA ($\chi_0$) --- shows no significant frequency dependence. Hence, the observed polarization of the burst is consistent with a frequency-independent GC trajectory and an $\rm RM = -8.85$ rad~m$^2$. This suggests that the LBM interpretation of the temporal variation of polarization state of \qp\ is self-consistent, although not necessarily unique. 

As described in Section~\ref{sec:newer}, the great circle trajectories traced by sub-bursts $A$ and $B$ of \newer\ have different orientations on the \pncr\ sphere. In the LBM interpretation of the great circle trajectories, this could be attributed to the difference between the properties of the LBMs associated with the two sub-bursts. The difference in $\chi_0$ could be explained by different orientations of the magnetic field with respect to the line of sight, which is possible if the two sub-bursts originate in different parts of the magnetosphere of a compact object. We note that \citet{twins} suggested a scenario where the two sub-bursts $A$ and $B$ of \newer\ are associated with emission from opposite magnetic poles of a compact magnetized progenitor, which could also lead to different orientations of the perpendicular component of the magnetic field associated with the two sub-bursts. The lack of significant frequency dependence of $\chi_0$ in \newer\ (see Figure~\ref{fig_210912_pspeca}) suggests that the average RM of $4.55$ rad m$^{-2}$ represents the ``true" degree of Faraday rotation for this FRB. 

\begin{figure*}[t!]
\centering
\includegraphics[width=0.49\linewidth]{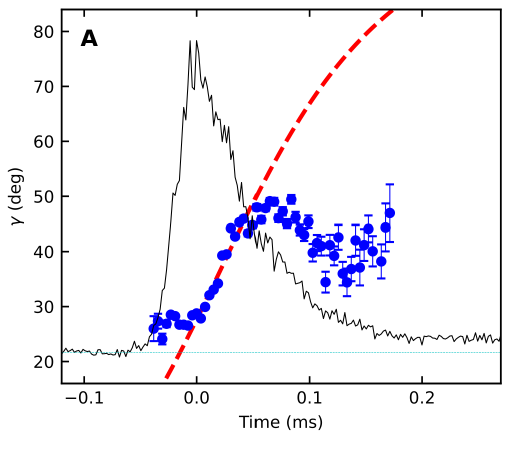}
\includegraphics[width=0.49\linewidth]{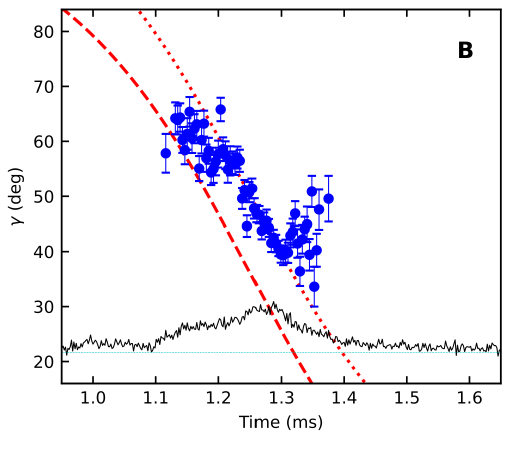}
\caption{\textbf{Variation of the inferred intrinsic PA ($\gamma = \Gamma - \phi$) across sub-burst $A$ [A, \textit{left panel}] and sub-burst $B$ [B, \textit{right panel}] of \newer.} Corrections for $\rm RM = 4.55\;rad\:m^2$ have been applied. The thick dashed lines represent the inferred RVM with $\alpha = 80.2 \dg$, $\Theta_A = 62.5\dg$ and $\beta_A = 17.7\dg$. The dotted line in the right panel shows the dashed line shifted by $14\dg$. See Section~\ref{sec:gama} for details.     
\vspace*{0.7cm}
\label{fig_newerpa}}
\end{figure*}

\subsection{Motion along the Great Circle} \label{sec:gama}

The motion along the great circle is parametrized by $\Gamma (t)$, with $2\Gamma$ representing the angular distance from the equator measured along the great circle (see Appendix~\ref{app:gcmod}). In the LBM interpretation, the observed time dependence of $\Gamma$ is related to the intrinsic PA ($\gamma$) variation of the emission as $\Gamma = \gamma - \phi$. Temporal variations of the PA of the linearly polarized emission have been observed in many other FRBs \citep[see e.g.][]{pandhi24chimepol,Sherman2023,Luo2020}, and are likely to be intrinsic to the emission source. 

Figure~\ref{fig_bothgama} shows the motion along the great circle --- i.e.\ the temporal evolution of $\Gamma$ --- across the primary sub-bursts of \qp\ and \newer\ in different frequency sub-bands. In \qp, the first peak does not show any frequency dependence of $\Gamma$, although the second peak exhibits measurable differences among the four sub-bands. The change in $\Gamma$ at the second peak --- with respect to its frequency independent value at the maximum of the first peak --- is larger at lower frequencies. We note that the first peak of \qp\ is almost fully linearly polarized and exhibits a roughly constant RM, while the second peak has a varying RM as well as significant circular polarization. The primary sub-burst of \newer\ exhibits frequency-dependent motion along the great circle, varying RM as well as significant circular polarization -- the same features as the second peak of \qp. Presence of circular polarization, apparent variation of RM and frequency dependence of $\Gamma$ thus appear to be connected, and have the same underlying reason. The observed frequency dependence of $\Gamma$ is not naturally explained by the LBM scenario. Under the assumption that $\phi$ is frequency independent, the observed frequency dependence of $\Gamma$ may be attributed to the frequency dependence of the intrinsic PA ($\gamma$) --- possibly due to propagation effects through a different medium before encountering the LBM (e.g. the magnetosphere of a compact source), or different frequencies being emitted from different regions and thus tracing differently orientated local magnetic fields.

In the LBM scenario, the intrinsic PA can be estimated as $\gamma = \Gamma - \phi$. Figure~\ref{fig_newerpa} shows the variation of $\gamma$ across sub-bursts $A$ and $B$ of \newer\ calculated using $\phi = -\chi_0$ for sub-burst $A$ and $\phi = 90\dg - \chi_0$ for sub-burst $B$. The two sub-bursts exhibit opposite sense of intrinsic PA variation. Following the method described in \citet{twins}, we estimated the rotating vector model \citep[RVM;][]{rad69rvm} parameters from the intrinsic PA variation, which are inclination $\alpha = 80.2\dg \pm 0.6\dg$, magnetic obliquity for the primary pole $\Theta_A = 62.5\dg \pm 0.5\dg$ and impact angle for the primary pole $\beta_A=17.7\dg \pm 0.5\dg$. As shown in Figure~\ref{fig_newerpa}, although the PA variation trend is consistent an RVM with these parmaters, there is an offset of $\approx 14\dg$ between the two sub-bursts. If the intrinsic PA variation in the two sub-bursts of \newer\ is indeed associated with the emission beams from opposite magnetic poles crossing the line of sight, this offset could arise from various complex magnetospheric effects which could also be responsible for the observed frequency dependence of $\Gamma$ \citep[see e.g.][]{hibschman01rvm,blaskiewicz91apj}.

\subsection{Apparent Rotation Measure Variation} \label{sec:rmvar}

The ``apparent" RM, estimated from the observed PA of linear polarization, can be attributed to a combination of the frequency dependence of $\alpha$, $\delta$ and $\Gamma$ which is in general complex. As described in the previous sections, $\delta$ does not show measurable frequency dependence for \qp\ or \newer\ within the observing band; however, a weak frequency dependence of $\delta$ may also lead to a significant ``effective RM'' \citep[see e.g.][]{lyutikov22apj}.  

In the LBM scenario, when circular polarization vanishes (i.e.\ $\Gamma=0$ or $\gamma=\phi$) the apparent RM represents the frequency dependence of $\alpha$ (which arises from Faraday rotation in the ISM/IGM between the LBM and the observer). This is broadly consistent with the observations for \qp\ and \newer\ described in Section~\ref{sec:gcfreq}. The temporal variation of $\Gamma$ leads to time-varying frequency dependence of the observed Stokes parameters which causes the gradual temporal variation of the apparent RM across the bursts.  

\vspace*{0.8cm}

\section{Discussion} \label{sec:discussion}

\subsection{Possibility of Instrumental Effects} \label{sec:instrument}

In general, instrumental polarization effects may lead to various systematic features on the observed polarization of radio emission. We note that the inclinations ($\Delta_0$) of the great circle trajectories associated with the primary sub-bursts of \qp\ and \newer\ are very similar (although the secondary sub-burst of \newer\ has a different inclination), which may point to an  instrumental origin of these trajectories. Here we investigate if the great circle trajectories of the polarization vector of \qp\ and \newer\ can originate from instrumental effects under two possible scenarios related to the origin of the observed circular polarization.

\textit{If these two FRBs have significant intrinsic circular polarization}: In this scenario the total Stokes V is the summation of the intrinsic Stokes V and the instrumental Stokes V originating from leakage. As the leakage is expected to be independent of the intrinsic circular polarization, it would have to be a remarkable coincidence that they add up to a total Stokes V which keeps the resultant polarization vector on a great circle trajectory on the \pncr\ sphere. 

\textit{If these two FRBs have zero/negligible intrinsic circular polarization}: In this scenario the observed Stokes V entirely originates from polarization leakage. The degree of polarization leakage ($\gtrsim50\%$) required to give rise to fractional circular polarization as high as what is observed in \qp\ and \newer\ is much larger than the typical leakage for ASKAP (which is only a few percent). Further discussion on the possibility of polarization leakage causing the observed trajectories is given in Appendix~\ref{app:poleakage}.

Both the scenarios described above hence cannot cause the observed great circle trajectories of the polarization vector for \qp\ and \newer.

\subsection{Possible Linear Birefringent Media} \label{sec:lbmorigin}

As discussed in Section~\ref{sec:lbm}, propagation through an LBM may result in a great circle trajectory of the polarization vector on the surface of the \pncr\ sphere. The observed difference in the great circle trajectories in the two sub-bursts of \newer\ (which are separated only by 1.3 ms) suggest that the LBM is located close to the emission source, and has different orientations of its axes in the two sub-bursts. A symmetric pair-plasma ($e^-$ \& $e^+$) in the presence of a magnetic field perpendicular to the direction of propagation have linear natural polarization modes and hence can act as an LBM \citep[e.g.][]{lyutikov22apj}. The orientation of the fast and slow axes are determined by that of the perpendicular component of the magnetic field on the sky plane, and the phase delay between the natural modes is given by
\begin{equation}
    \delta = \frac{2e^2}{m_e c^2}\: n_e\:\Delta l \: \lambda
\end{equation}
in the strong magnetic field regime (cyclotron frequency much larger than the frequency of the radiation), where $e$, $m_e$ and $n_e$ are the charge and the mass and the number density of electrons and $\Delta l$ is the thickness of the plasma shell. We note that, in this scenario, the magnetic field associated with this plasma has to be predominantly perpendicular to the line of sight for the natural polarization modes to be approximately linear. Such an orientation of the magnetic field naturally arises in the outer magnetosphere or the near-wind region of a neutron star \citep{lyutikov22apj}. In case of \newer, if the two sub-bursts are associated with emission from different regions of a compact object (e.g. a neutron star), different orientations of the local magnetic field (and hence the axes of the LBM) in the outer magnetosphere (or the near-wind region) would naturally explain the observed difference in the \pncr\ sphere trajectories for the two sub-bursts. This scenario is consistent with a magnetospheric origin of FRBs, supporting the conclusion of some recent studies \citep[e.g.][]{niu24pajump,nimmo25mag}.

Assuming a pair plasma in a high magnetic field such that the cyclotron frequency is much higher than the frequency of the EM waves, we find that a plasma screen with a column density of $\sim 10^{10}-10^{11}$ cm$^{-2}$ may lead to the observed inclinations ($\Delta_0$) of the great circle trajectories on the \pncr\ sphere. This column density is not unreasonable for a relatively thin plasma screen near the light cylinder of a neutron star \citep[e.g.][]{beloborodov07mag,thompson02mag}. As the phase difference between the two natural polarization modes varies linearly with the wavelength of the EM radiation in this scenario, this is also consistent with the lack of significant frequency dependence of $\Delta_0$ within the observing frequency band.

As mentioned previously, no further bursts have been reported from either \newer\ or \qp\ to date. However, detection of more bursts from these sources in future could potentially provide an excellent opportunity to test the hypothesis described in this section and to probe the magnetic field properties of the emitting objects.

\subsection{Other FRBs} \label{sec:others}

In the current sample of CRAFT detected FRBs \citep{shannon24craftics}, no other FRB is found to exhibit great circle trajectories of the polarization vector on the \pncr\ sphere. Incidentally, \older, which shows remarkable similarities with \newer\ --- including similar burst profiles and apparent RM variation, although a lower fraction of circular polarization \citep{twins} --- does not show a similar trajectory of the polarization vector on the \pncr\ sphere. The polarization vector associated with the primary sub-burst of \older\ traces a small closed loop on the surface of the \pncr\ sphere \citep{cho20apjl}. 

Even if this phenomenon is ubiquitous in FRBs, in order to observe the great circle trajectories some additional criteria need to be satisfied. Firstly, these effects would not be observable in strongly scattered FRBs as scattering mixes time-dependence with frequency-dependence and alters the temporal variation of polarization state. Secondly, these effects are likely to be observable only in FRBs with relatively simple burst morphology. Complex burst structure with multiple overlapping components --- presumably associated with different emitting regions and thus having independent polarization behaviour --- would make it difficult to identify such effects. Thirdly, sufficient temporal variation of the intrinsic PA during the burst is required for the polarization vector to trace a significant arc along a great circle. With the current sample of CRAFT FRBs with high-time-resolution polarization data ($\approx$ 30), it is thus unsurprising that this effect has only been observed in two FRBs. A systematic study of polarization for a large sample of FRBs with full polarization data is necessary to determine how common these observed effects are.  

\section{Summary} \label{sec:summary}

We investigated temporal variation of the polarization states of two fast radio bursts, \qp\ and \newer, which show time-dependent conversion between linear and circular polarizations. The polarization vector was found to trace great circle trajectories on the surface of the \pncr\ sphere for both FRBs, with the two sub-bursts of \newer\ tracing different great circles. The trajectories of the polarization vector show weak frequency dependence within the observing frequency band. The observed temporal variation of the polarization vector for these two FRBs are unlikely to have originated from instrumental effects. These polarization features may be signatures of a transition between two partially coherent orthogonal polarization modes or propagation through a birefringent medium. However, we find that transition between two orthogonal polarization modes or effects of propagation through an elliptically birefringent medium are unlikely to give rise to the observed polarization behaviour. The observed variations of the polarization states of these two FRBs are qualitatively consistent with a magnetospheric origin of the bursts and the effects of propagation through a birefringent medium with linearly polarized modes in the outer magnetosphere or near-wind region of a neutron star.

\section*{Acknowledgements}

We thank Don Melrose for important suggestions that led to the interpretations of the observations presented in this paper. AB thanks Dipanjan Mitra, Adrian Sutinjo, Marcin Sokolowski and Marcus E. Lower for useful comments and discussion regarding this work. We acknowledge the custodians of the land this research was conducted on, the Whadjuk (Perth region) Noongar people and pay our respects to elders past, present and emerging. CWJ and MG acknowledge support by the Australian Government through the Australian Research Council's Discovery Projects funding scheme (project DP210102103). RMS acknowledges support through ARC Future Fellowship FT190100155. RMS and ATD acknowledge support by the Australian Government through the Australian Research Council’s Discovery Projects funding scheme (project DP220102305). This scientific work uses data obtained from the Australian Square Kilometre Array Pathfinder (ASKAP), located at Inyarrimanha Ilgari Bundara / the Murchison Radio-astronomy Observatory. We acknowledge the Wajarri Yamaji People as the Traditional Owners and native title holders of the Observatory site. ASKAP uses the resources of the Pawsey Supercomputing Research Centre. CSIRO’s ASKAP radio telescope is part of the Australia Telescope National Facility. Establishment of ASKAP, Inyarrimanha Ilgari Bundara, the CSIRO Murchison Radio-astronomy Observatory and the Pawsey Supercomputing Research Centre are initiatives of the Australian Government, with support from the Government of Western Australia and the Science and Industry Endowment Fund. Operation of ASKAP is funded by the Australian Government with support from the National Collaborative Research Infrastructure Strategy. This work was performed on the OzSTAR national facility at Swinburne University of Technology. The OzSTAR program receives funding in part from the Astronomy National Collaborative Research Infrastructure Strategy (NCRIS) allocation provided by the Australian Government, and from the Victorian Higher Education State Investment Fund (VHESIF) provided by the Victorian Government. The National Radio Astronomy Observatory is a facility of the National Science Foundation operated under cooperative agreement by Associated Universities, Inc. This research has made use of NASA's Astrophysics Data System Bibliographic Services.

\facilities{ASKAP}

\software{
\textsc{Matplotlib} \citep{Matplotlib2007}, \textsc{NumPy} \citep{Numpy2011}, \textsc{SciPy} \citep{SciPy2019}, \textsc{AstroPy} \citep{astropy:2022}
}

\appendix 

\renewcommand\thefigure{A\arabic{figure}}
\setcounter{figure}{0} 
\renewcommand\thetable{A\arabic{table}}
\setcounter{table}{0} 

\section{Variations in Position and Ellipticity Angles Due to Elliptical Birefringence} \label{app:paeagfr}

The changes in the PA and EA with $\eta$ due to elliptical birefringence (described in Section~\ref{sec:ebm}) can be determined by a coordinate transformation from the reference frame of the GC plane to the reference frame of the Poincar\'e sphere. For the purpose of the transformation, the GC reference frame is defined as $x'$-$y'$-$z'$ with the GC located in the $y'$-$z'$ plane. Since the radius of the Poincar\'e sphere and GC are equal to 1, the equation for the GC as a function of $\eta$ is $\cos^2\eta+\sin^2\eta=1$. The Q-U-V vector resulting from the coordinate transformation can then be used to calculate the PA and EA from their standard definitions.
\begin{equation}
\chi(\eta) = \frac{1}{2}\arctan{\left[-\frac{\cos\Delta\tan\eta + \tan(2\chi_o)}
             {\cos\Delta\tan\eta\tan(2\chi_o) - 1}\right]}
\label{eqn:PA}
\end{equation}
\begin{equation}
\psi(\eta) = \frac{1}{2}\arctan{\left[-\frac{\sin\Delta\tan\eta}
       {\sqrt{1+\cos^2\Delta\tan^2\eta}}\right]}
\label{eqn:EA}
\end{equation}

\section{Derivation of Great Circle Trajectories due to Linear Birefringence} \label{app:gcmod}

Using Jones vector representation in an arbitrarily chosen linear basis, the electric field at the emission site is given by
\begin{equation*}
    \Vec{E} = E_0 \begin{bmatrix}
        \cos{\gamma (t)} \\
        \sin{\gamma (t)}
    \end{bmatrix},
\end{equation*}
where $\gamma$ is the angle between the electric field on the sky plane and the reference axis. An intrinsic time dependence of $\gamma$ can be attributed to the properties of the emission source (see Section~\ref{sec:gama}). An LBM introduces a phase difference $\delta$ between the two orthogonal linear components of the electric field along its optical axes (i.e.\ the `fast' axis and the `slow' axis). In our arbitrarily chosen coordinate system, the Jones matrix of an LBM is given by 
\begin{equation*}
    R_{\delta, \phi}  = 
    \begin{bmatrix}
        \cos{\phi} & -\sin{\phi} \\
        \sin{\phi} & \cos{\phi}
    \end{bmatrix}
    \begin{bmatrix}
        e^{-i\delta /2} & 0 \\
        0 & e^{i\delta /2}
    \end{bmatrix}
    \begin{bmatrix}
        \cos{\phi} & \sin{\phi} \\
        -\sin{\phi} & \cos{\phi}
    \end{bmatrix},
\end{equation*}
where $\phi$ is the angle between the reference axis and the fast axis of the LBM. $R_{\delta , \phi}$ incorporates transformation to the LBM coordinate system (defined by its fast and slow axes), application of the phase delay and transformation back to the initial coordinate system. 

Effects of Faraday Rotation of the PA of linear polarization in the magnetized interstellar and intergalactic media is represented by the Jones matrix
\begin{equation*}
    F_{\alpha}  = \begin{bmatrix}
        \cos{\alpha (\lambda)} & \sin{\alpha (\lambda)} \\
        -\sin{\alpha (\lambda)} & \cos{\alpha (\lambda)}
    \end{bmatrix},
\end{equation*}
where $\alpha (\lambda)$ is the frequency dependent rotation angle. Assuming that the radio emission travelled through an LBM and the ISM/IGM, the observed electric field can be written as
\begin{align*}
    \Vec{E}_{obs} & = F_{\alpha} \; R_{\delta, \phi} \; \vec{E} \\
    & = E_0 \begin{bmatrix}
        \cos{\frac{\delta}{2}}\:\cos{(\gamma - \alpha)} - i \: \sin{\frac{\delta}{2}}\:\cos{(\gamma + \alpha - 2\phi)} \\
        \cos{\frac{\delta}{2}}\:\sin{(\gamma - \alpha)} + i \: \sin{\frac{\delta}{2}}\:\sin{(\gamma + \alpha - 2\phi)} 
    \end{bmatrix}
\end{align*}
and the observed Stokes parameters are given by 
\begin{equation}
    \begin{bmatrix}
        Q \\ U \\ V
    \end{bmatrix}
    = E_0^2 \begin{bmatrix}
        - \cos^2{\frac{\delta}{2}} \: \cos{(2\gamma - 2\alpha)} - \sin^2{\frac{\delta}{2}} \: \cos{(2\gamma + 2\alpha - 4 \phi)} \\ 
        \cos^2{\frac{\delta}{2}} \: \sin{(2\gamma - 2\alpha)} - \sin^2{\frac{\delta}{2}} \: \sin{(2\gamma + 2\alpha - 4 \phi)} \\ 
        \sin{\delta}\: \sin{(2\gamma - 2\phi)}
    \end{bmatrix}
\end{equation}
in the convention described in Section~\ref{sec:methods}. 

\begin{figure*}[t]
\centering
\includegraphics[scale=0.9,trim={0.1cm 0cm 0.15cm 0cm},clip]{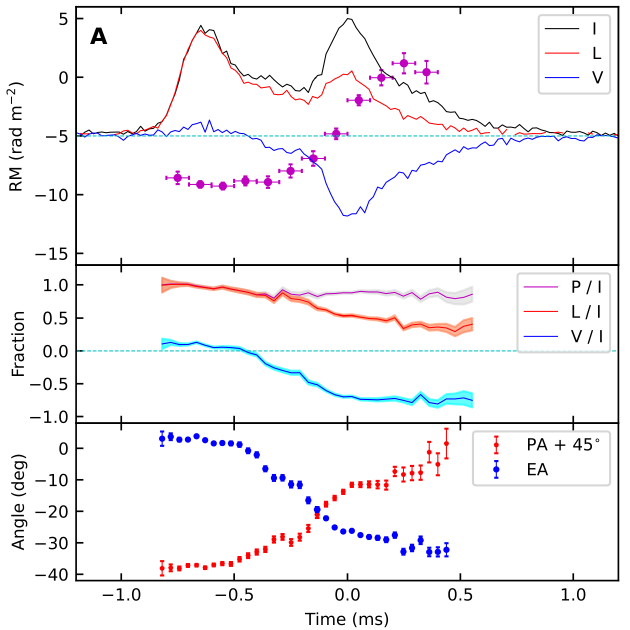}
\includegraphics[scale=1.1,trim={0.0cm 0cm 0cm 0cm},clip]{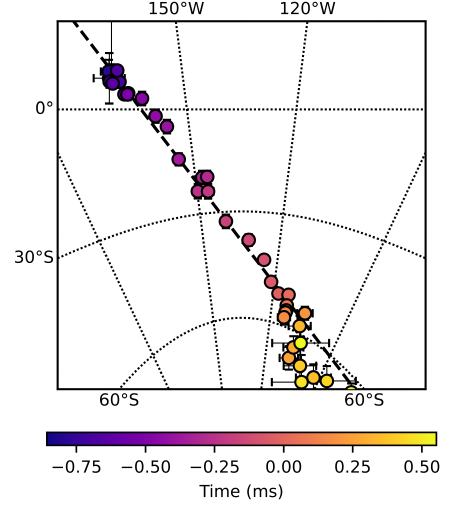}
\caption{\textbf{Frequency averaged polarization time profile of \qp\ in the primary sub-burst [A, \textit{left}] and in the fainter trailing sub-bursts [B, \textit{right}] after correcting for $\rm RM = -8.85\;rad\:m^2$.} The time axis has different scales in the left and right panels.  See Figure~\ref{fig_230708_fullpol575} for a description of the symbols. 
\label{fig_230708_fullpol885}}
\end{figure*}

Rotating the $Q-U$ plane (about the V-axis) by an angle of $2\zeta$, the transformed Stokes parameters in the new coordinate system are given by
\begin{equation}
    \begin{bmatrix}
		Q' \\
		U'
	\end{bmatrix}
	= 
	\begin{bmatrix}
		\cos{2\zeta} & \sin{2\zeta} \\
		-\sin{2\zeta} & \cos{2\zeta} 
	\end{bmatrix}    				
	\begin{bmatrix}
		Q  \\
    		U 
	\end{bmatrix} 
	=
	\begin{bmatrix}
        - \cos^2{\frac{\delta}{2}} \: \cos{(2\zeta + 2\gamma - 2\alpha)} - \sin^2{\frac{\delta}{2}} \: \cos{(2\zeta - 2\gamma - 2\alpha + 4 \phi)} \\ 
        \cos^2{\frac{\delta}{2}} \: \sin{(2\zeta + 2\gamma - 2\alpha)} + \sin^2{\frac{\delta}{2}} \: \sin{(2\zeta - 2\gamma - 2\alpha + 4 \phi)} 
    \end{bmatrix}.
\end{equation} 
$V$, $L$ and $\psi$ remain unchanged, while the PA of linear polarization transforms as 
\begin{equation}
\chi' = \chi - \zeta.
\end{equation} 
In the rotated system, the Stokes parameters follow the relation
\begin{equation}
\frac{\tan{(2\psi')}}{\sin{(2\chi')}} = \frac{\tan{(2\psi)}}{\sin{(2\chi')}} = \frac{V'}{U'} = \frac{V}{U'} = \frac{\sin{\delta} \sin{(2\gamma - 2\phi)}}{\cos^2{\frac{\delta}{2}} \sin{(2\gamma - 2\alpha + 2\zeta)} - \sin^2{\frac{\delta}{2}} \sin{(2\gamma + 2\alpha - 2\zeta - 4\phi)}},
\end{equation}
which reduces to a great circle when
\begin{equation}
\zeta \equiv \alpha - \phi,
\end{equation}
and the equation of the great circle, in the original (un-rotated) coordinates, is given by
\begin{equation}
\frac{\tan{(2\psi')}}{\sin{(2\chi')}}  = \frac{\tan{(2\psi)}}{\sin{(2\chi - 2\zeta)}} = \tan{\delta}.
\end{equation}
The motion along a great circle trajectory is parametrized by 
\begin{equation}
\Gamma = \frac{1}{2} \arctan{\left( \frac{\sqrt{U'^2 + V^2}}{-Q'} \right)} = \gamma - \phi
\end{equation}
where $2\Gamma$ is the angular separation from the equator along the great circle. The equation of the great circle can also be written as 
\begin{equation}
    \tan{2\psi} = \tan{\delta} \: \sin{(2\chi - 2\alpha + 2\phi)} 
\label{eq:lbmgc2}
\end{equation}
which is the same as Equation~\ref{eqn_mc} with $(\delta, \alpha - \phi)$ $= (\Delta_0 +n\pi, \chi_0$) or $(n\pi - \Delta_0, \chi_0 \pm \pi/2)$ for arbitrary integral values of $n$.

\begin{figure*}[t]
\centering
\includegraphics[scale=0.71,trim={0.0cm 0cm 0.2cm 0cm},clip]{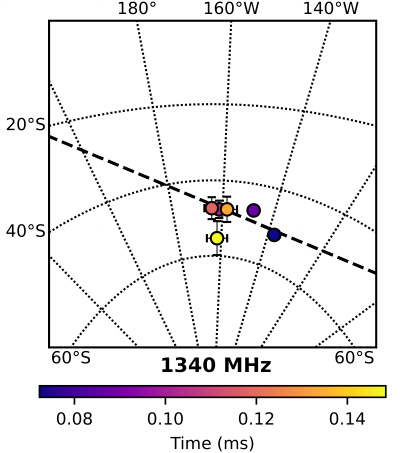}
\includegraphics[scale=0.71,trim={0.5cm 0cm 0.3cm 0cm},clip]{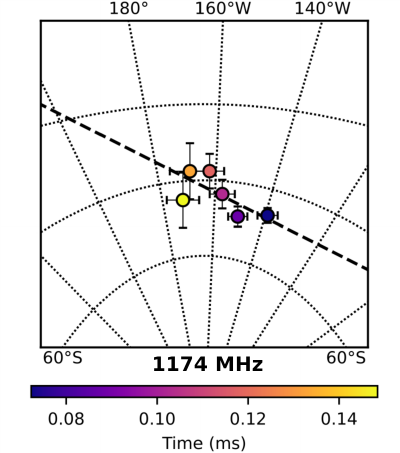}
\includegraphics[scale=0.71,trim={0.0cm 0cm 0.2cm 0cm},clip]{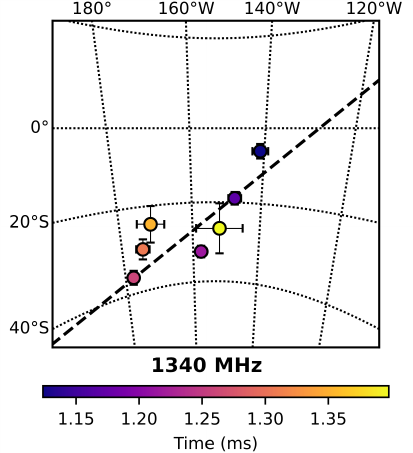}
\includegraphics[scale=0.71,trim={0.6cm 0cm 0.2cm 0cm},clip]{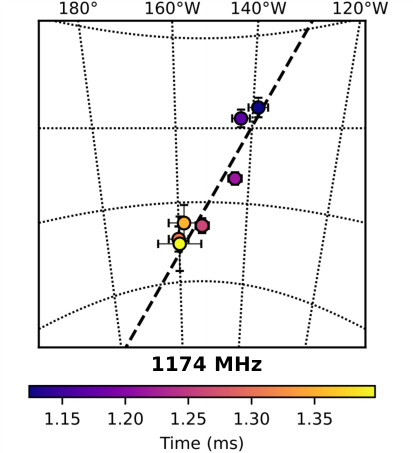}
\caption{\textbf{Temporal evolution of polarization state across the `tail' of sub-burst $A$ [A,B, \textit{first two panels}] and across sub-burst $B$ [C,D, \textit{last two panels}] of \newer\ on the \pncr\ sphere after correcting for $\rm RM = 4.55\;rad\:m^2$ and averaging over two frequency sub-bands centred at 1339.8 MHz and 1174.4 MHz, respectively.} Each data point represents the polarization state of radio emission averaged over a 15 $\mu$s time bin in the first two panels and 46 $\mu$s time bin in the last two panels. The best-fit great circle is shown as a dashed black line in each panel. 
\label{fig_210912_atlb_pspsb_455}}
\end{figure*}

\begin{figure}[t]
\centering
\includegraphics[width=0.49\linewidth]{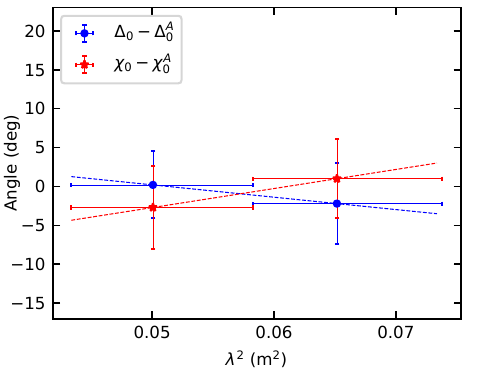}
\includegraphics[width=0.49\linewidth]{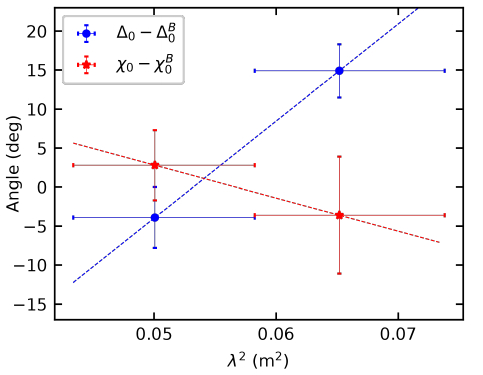}
\caption{\textbf{Wavelength dependence of the best-fit great circle parameters ($\Delta_0$, $\chi_0$) in \newer\ after correcting for $\rm RM = 4.55\;rad\:m^2$ in the tail of sub-burst $A$ [A, \textit{left}] and in sub-burst $B$ [B, \textit{right}], respectively.} Best-fit parameters for the full-band have been subtracted from these measurements. The dashed lines show best-fit linear variations of the corresponding angles with $\lambda^2$.  
\label{fig_210912_pspecb}}
\end{figure}

\section{Polarization Profiles of \qp\ after correcting for RM = -8.85 rad m$^{-2}$} \label{app:qp885}

Figure~\ref{fig_230708_fullpol885} shows the temporal variation of polarization properties of \qp\ after correcting for $\rm RM = -8.85$ rad m$^{-2}$. As discussed in Section~\ref{sec:gcfreq}, the results do not show any significant difference from those obtained after correcting for $\rm RM = -5.75$ rad m$^{-2}$ (shown in Figure~\ref{fig_230708_fullpol575}).

\section{Frequency dependence of great circle trajectories for \newer\ } \label{app:newersubband}

Frequency dependence of the great circle trajectory associated with the peak of sub-burst $A$ has been described in Section~\ref{sec:freqdepnewer}. Figure~\ref{fig_210912_atlb_pspsb_455} shows the great circle fits to the trajectories of the polarization vector for the tail of sub-burst $A$ and for sub-burst $B$ in two frequency sub-bands. As shown in Figure~\ref{fig_210912_pspecb}, the great circle parameters have large uncertainties and hence their frequency dependence is poorly constrained.

\section{Further Discussion on Polarization Leakage} \label{app:poleakage}

A polarization calibrator (the Vela pulsar) was observed within a few hours of detection of both \qp\ and \newer, and was used for polarization calibration. No evidence of high polarization leakage (more than a few percent) was found for the calibrator. Polarization calibration solutions obtained for other CRAFT detected FRBs, using the same polarization calibrator, are also consistent with a few percent of polarization leakage. A temporary rise in the leakage amplitudes --- when these two FRBs were detected --- is not likely to cause the same inclination of the great circle trajectories for \qp\ and \newer. Both these FRBs were detected in different observation bands (centered at different frequencies) and more than two years apart. Additionally, the two sub-bursts of \newer\ were observed almost simultaneously in the raw (un-dedispersed) frame. Therefore, variable instrumental leakage should affect both sub-bursts equally, which is not what is observed.  

\bibliography{frbref}{}
\bibliographystyle{aasjournal}

\end{document}